\documentclass[preprint,aps,showpacs]{revtex4}

\usepackage{epsfig}
\usepackage{dcolumn}
\usepackage{bm}

\begin{document}

\title{Phase behavior and dynamics of a micelle-forming triblock copolymer system}

\author{P. Harsha Mohan}

\author{Ranjini Bandyopadhyay}
 \email{ranjini@rri.res.in}
\affiliation{Raman Research Institute, Bangalore 560080, INDIA}
%


\date{\today}

\begin{abstract}
Synperonic F-108 (generic name, 'pluronic') is a micelle forming triblock copolymer of type ABA, where A is polyethylene oxide (PEO) and B is polypropylene oxide (PPO). At high temperatures, the hydrophobicity of the PPO chains increase, and the pluronic molecules, when dissolved in an aqueous medium, self-associate into spherical micelles with dense PPO cores and hydrated PEO coronas. At appropriately high concentrations, these micelles arrange in a face centred cubic lattice to show inverse crystallization, with the samples exhibiting high-temperature crystalline and low-temperature fluidlike phases. By studying the evolution of the elastic and viscous moduli as temperature is increased at a fixed rate, we construct the concentration-temperature phase diagram of Synperonic F-108. For a certain range of temperatures and at appropriate sample concentrations, we observe a predominantly elastic response. Oscillatory strain amplitude sweep measurements on these samples show pronounced peaks in the loss moduli, a typical feature of soft solids. The soft solid-like nature of these materials is further demonstrated by measuring their frequency dependent mechanical moduli. The storage moduli are significantly larger than the loss moduli and are almost independent of the applied angular frequency. Finally, we perform strain rate frequency superposition (SRFS) experiments to measure the slow relaxation dynamics of this soft solid. 

\end{abstract}

\pacs{36.20.-r, 83.80.Uv, 83.50.-v, 83.80.Qr} 
\maketitle

\section{Introduction}

Amphiphilic, or associative, polymers contain hydrophobic or
hydrophilic groups that can aggregate to form micellar structures
similar to those formed by simple surfactant molecules
\cite{semenov_macromols}. Examples of associative polymers include
telechelic polymer chains, characterized by strongly adsorbing groups
\cite{milner_macromols}, and micelle-forming block copolymers
\cite{block_applied}.  

When in solution, block copolymers can exhibit very rich phase
behavior. Such phase behavior depends strongly on the combination of
monomers constituting the block, molecular architecture, composition,
and molecular size, and is very sensitive to changes in the
interaction parameters between the different polymer units and the
solvent \cite{bates_science}. Block copolymers can aggregate to form
structures as diverse as bicontinuous phases, microemulsions and
micelles \cite{bates_science,chen_book}. These materials find
  important uses as non-ionic surfactants in industrial and technical
  applications \cite{alexand_eng} and as novel polymer therapeutics
  for drug and gene delivery \cite{kabanov_release}.  Copolymers that consist of a central polypropylene oxide (PPO) block,
  and with polyethylene oxide (PEO) blocks on either sides, is one such
  example of a triblock copolymer system. 
The PPO block, which is predominantly hydrophilic at low temperatures, becomes increasingly hydrophobic as the temperature is raised \cite{israel_review}. A direct consequence of this behavior is the self-association of the triblock copolymers into spherical micelles above a critical polymer concentration and a critical temperature, with the PPO block forming a dense core and the hydrated PEO chains forming a swollen corona \cite{zhou_jcis,wanka_polsci,brown_jpc,mortensen_epl,yardimci_jcp}. Systematic studies using x-ray diffraction (XRD) and small angle neutron scattering (SANS) show that while solutions of triblock copolymers form micellar liquids at low temperatures \cite{goldmints_lang}, at appropriately high temperatures and above certain concentrations, these micelles can arrange in domains characterized by cubic crystalline order \cite{yardimci_jcp,mortensen_prl,prudhomme_langmuir}. Changes in the aggregation behavior of individual micelles and modifications of the intermicellar interactions with changes in temperature and polymer concentration have also been investigated extensively using SANS, small angle x-ray scattering (SAXS) and light scattering \cite{goldmints_lang2,lobry_pre,castelletto_sm}. Studies on the influence of the effective potential between spherical micelles on the lattice ordering have also been performed \cite{hamley_lang}. Simultaneous SANS and rheology experiments to study the structure-viscoelasticity correlations in the face-centred cubic crystalline phase formed by the long-chain triblock copolymer system F-127 point to the repulsion between close-packed spherical micelles as the dominant mechanism for the formation of the observed soft solid-like phase \cite{prudhomme_langmuir}. Softer intermicellar interaction potentials in the copolymer system L-64, on the other hand, gives rise to the ordering of the spherical micelles in a body centred cubic lattice \cite{lobry_pre}. 

While most of the earlier work on block copolymers melts and solutions exclusively focused on their structure and phase behavior, experimental studies of their dynamics and rheology have been reported more recently \cite{yardimci_jcp,prudhomme_langmuir,lau_jpol}. Intermediate scattering functions measured in neutron spin echo (NSE) experiments on concentrated triblock copolymer solutions show the existence of two distinct relaxation regimes \cite{yardimci_jcp}. The faster relaxation is identified with the polymer brush-like longitudinal diffusive modes in the PEO corona, while the wave vector dependence of the slower relaxation process is attributed to the Rouse modes of the PPO chains constituting the concentrated micellar corona. Correlations between the nonlinear viscoelasticity of these soft micellar crystals and their structure have also been studied using rheology and {\it in situ} SAXS \cite{molino_epjb,eiser_epje}. These studies find that at the lowest shear rates, the flow is defect-mediated, while at intermediate shear rates, a layer sliding mechanism dominates. At high shear rates, severe shear-thinning accompanied by the melting of the polycrystalline domains is observed. 

In this paper, we perform oscillatory rheology experiments to understand the flow behavior and the relaxation dynamics of aqueous solutions of the micelle-forming triblock copolymer system F-108. We perform large amplitude oscillatory strain (LAOS) and frequency sweep experiments to show that the system forms a soft solid over a large range of temperatures and copolymer concentrations. Previous SANS experiments performed to elucidate the phase behavior of this system show that at appropriate copolymer concentrations and with increasing temperatures, the system shows a sequence of phase transformations from individual copolymer chains dissolved in solution, to micelles characterized by liquid-like order, to soft cubic crystals and finally back to the micellar liquid phase \cite{yardimci_jcp}. We perform temperature sweep and frequency response experiments over a wide range of copolymer concentrations and show the presence of an intermediate `soft solid' phase similar to that observed in the diblock copolymer E$_{87}$B$_{18}$ \cite{castelletto_jcp}. This new phase is characterized by an elastic modulus that is significantly higher than the viscous modulus, but which is at least two orders of magnitude lower than the elastic modulus measured in the cubic crystalline phase. Finally, we perform strain rate frequency superposition (SRFS) experiments \cite{wyss_prl} to calculate the slow relaxation dynamics of the system and compare these results with an average relaxation time extracted by fitting our frequency response data to the Cole Davidson model \cite{cole_jcp,davidson_jcp}.

\section{Experimental Methods}
\subsection{Materials and sample preparation}  
Synperonic F-108 (PEO$_{127}$PPO$_{48}$PEO$_{127}$) is purchased from Sigma Aldrich Ltd. and used as received without further purification. Each individual copolymer contains 80 wt.\% PEO and has a molecular weight of 14600 g/mol. Samples of several concentrations were made by dissolving appropriate quantities of the triblock copolymer in deionized and distilled water. The samples were stored for three days at temperatures below 5$^{\circ}$C, which is well below the critical micelle temperature, to ensure homogeneous mixing and equilibration. 
\subsection{Rheological measurements} 
Rheological data was acquired in an Anton Paar Modular Compact Rheometer, model number MCR 501, which is equipped with TruGap and Toolmaster to detect and control the measuring gap in the geometry, and which can perform direct strain oscillation measurements \cite{lauger_rheolacta}. A double gap geometry (DG26.7/Q1) capable of measuring stresses between 1.09 $\times$ 10$^{-3}$ Pa and 2.51 $\times$ 10$^{3}$ Pa, shear rates between 3.07 $\times$ 10$^{-6}$ /sec and 9.21 $\times$ 10$^{3}$ /sec, and a minimum strain of 2.93 $\times$ 10$^{-4}$ was used as the sample cell to perform the oscillatory measurements reported below. The temperature of the sample was controlled by means of a water circulation unit Viscotherm VT2 capable of controlling temperatures between 5$^{\circ}$C and 80$^{\circ}$C when the circulating fluid is water, and to much lower and higher temperatures when the circulating fluid is changed to acetone and silicon oil respectively. The sample was loaded at 5$^{\circ}$C to ensure a fluidlike consistency and its temperature was increased to the desired value at a controlled rate of 0.25 $^{\circ}$C/minute. Care was taken during each measurement to ensure that the torques, deformations and speeds recorded were well within the capabilities of the instrument as quoted by the manufacturer.

\section{Results and Discussions}
\subsection{Amplitude sweep measurements: observation of soft solid-like behavior} 
Complex fluids such as colloidal suspensions, surfactant solutions, polyelectrolytes and emulsions show very interesting linear and nonlinear  rheology \cite{larson_book}. The rheological response of these materials can be characterized by their complex shear modulus G$^{\star}(\omega)$ = G$^{\prime}(\omega)$ + {\it i}G$^{\prime\prime}(\omega)$ where G$^{\prime}$ is the elastic modulus, G$^{\prime\prime}$ is the viscous modulus and $\omega$ is the applied angular frequency \cite{macosko}. It is also useful to perform amplitude sweep experiments to isolate the linear viscoelastic regime of a material and to understand its response to strain deformations. Experiments with some polymeric systems point to four distinct behaviors of the large amplitude oscillatory strain (LAOS) response \cite{kyu_jnnfn}. These distinct behaviors depend strongly on the nature of the shear-induced structures formed and can be categorized in the following groups: strain thinning, strain hardening, weak strain overshoot and strong strain overshoot. 

For F-108 samples prepared at various concentrations, we performed strain amplitude sweep measurements by measuring the G$^{\prime}$ and G$^{\prime\prime}$ while varying the strain amplitude between 0.1\% and 100\% at a fixed angular frequency of 1 rad/sec. Fig. 1(a) shows our amplitude sweep data for a sample of concentration 0.25 g/cc acquired at a temperature of 60$^{\circ}$C. We note that the elastic and viscous moduli (G$^{\prime}$ and G$^{\prime\prime}$ are denoted by squares and circles respectively) do not quite exhibit a true linear regime over the range of strains explored. Above a critical yield strain, G$^{\prime}$ is found to decrease with strain $\gamma_{\circ}$ according to a power law relation of the type: G$^{\prime}(\gamma_{\circ}) \sim \gamma_{\circ}^{-\nu^{\prime}}$, while G$^{\prime\prime}$ decays according to the power law G$^{\prime\prime}(\gamma_{\circ}) \sim \gamma_{\circ}^{-\nu^{\prime\prime}}$. Fits to the power law decays of the moduli are shown by solid lines in the figure, and yield exponents $\nu^{\prime} \sim$ 1.24$\pm$0.06 and $\nu^{\prime\prime} \sim$ 0.82$\pm$0.06. The ratio of the exponents ${\nu^{\prime}}/{\nu^{\prime\prime}}$ is approximately 1.5. We would like to note here that in the case of a Maxwellian fluid which is characterized by a single relaxation time, ${\nu^{\prime}}/{\nu^{\prime\prime}}$ is expected to be 2 \cite{miyazaki_epl}. We believe that the existence of several closely spaced dynamical time scales in the samples studied by us accounts for this departure that we observe from the results expected for a Maxwell fluid.

A remarkable feature of the data for G$^{\prime\prime}$ is the appearance of a prominent peak before the power law decay, an ubiquitous feature in a diverse array of materials that can be classified as soft solids \cite{wyss_prl,miyazaki_epl}. Emulsions, pastes, concentrated colloidal suspensions, polymer gels \cite{bonn_science} and colloid-liquid crystal composites \cite{meeker_pre} exhibit several such unusual features in their rheology and are a few examples of these soft solids. Some of these soft solids can be categorized as `soft glassy' materials and are characterized by disordered and metastable structures \cite{sollich_prl}.

Figs. 1(b) and 1(c) show the strain responses of G$^{\prime}$ (squares) and G$^{\prime\prime}$ (circles) respectively at 60$^{\circ}$C for F-108 samples of different concentrations ranging between 0.16 and 0.29 g/cc. Remarkably, the exponents characterizing the power law decays of the viscoelastic moduli at large amplitude oscillatory strains are found to be independent of copolymer concentration. Another noteworthy feature is the appearance of the peak in G$^{\prime\prime}$ for all sample concentrations investigated. Such observations, that can be explained using mode coupling theory arguments \cite{miyazaki_epl}, clearly indicate that the universality of the strain response of F-108 extends even to its nonlinear regime. 
\subsection{Temperature and frequency sweep measurements: the
  concentration-temperature phase diagram of F-108} 

In order to study the temperature-dependence of the elastic and viscous moduli of F-108, and to use this information to construct a phase diagram, we measure G$^{\prime}$ and G$^{\prime\prime}$ as we vary the sample temperature in the double gap geometry at a constant rate of 0.25$^{\circ}$C/minute, while maintaining the angular frequency and strain amplitude at fixed values of 1 rad/sec and 0.5\% respectively. In Fig. 2, we plot the results of a typical temperature sweep run for a sample of concentration 0.28 g/cc. Clearly, the data can be divided into three distinct regimes: at the lowest temperatures (below 20.1$^{\circ}$C), G$^{\prime}$ (solid squares) is approximately equal to zero while G$^{\prime\prime}$ (open circles) has a small but measurable value. This indicates that the sample has a predominantly liquid-like response in this regime. At around 20.2$^{\circ}$C, a
finite G$^{\prime}$ appears, and both G$^{\prime}$ and G$^{\prime\prime}$ show an abrupt increase to a regime where the moduli are characterized by values of the order of 100 and 10 Pa respectively. Finally, there is another abrupt increase of both moduli at a temperature close to 30$^{\circ}$C to a third regime where G$^{\prime}$ and G$^{\prime\prime}$ saturate to values of 10000 and 1000 Pa respectively. Such a sequence of transitions of G$^{\prime}$ and G$^{\prime\prime}$ with increasing temperature is seen at all sample concentrations above 0.2 g/cc and has been summarized in the
inset of fig. 2. The temperature at which the transitions occur are found to shift to higher values as the sample concentration is decreased, such that for the lowest sample concentrations investigated, the second abrupt jump in the moduli is not observed. 

In order to construct a concentration-temperature phase diagram of Synperonic F-108, we perform a systematic study of the transitions of G$^{\prime}$ and G$^{\prime\prime}$ with temperature between these three regimes, for several sample concentrations. We identify the abrupt jumps of the mechanical moduli at certain temperatures as indicative of transformations between distinct phases, and use this information to plot the phase boundaries. This information, coupled with frequency response measurements that shed light on the mechanical properties of F-108, is used to understand the phase behavior of the samples. Our phase diagram is plotted in figure 3. At the lowest temperatures, throughout the range of concentrations explored, the
sample is in regime 1 (filled circles) where it is predominantly liquid-like. The liquid-like behavior in this regime points to the coexistence of unimers and micelles in solution under these experimental conditions, and is obvious from the frequency response measurement for a sample whose concentration (0.16 g/cc) and temperature (25$^{\circ}$C) values lie within this regime. This data, plotted in inset (a), shows that G${^{\prime}}(\omega) \sim \omega^{2}$ and G$^{{\prime\prime}}(\omega) \sim \omega$, which is indicative of a liquid-like response. The frequency response of the sample in the second regime, which is designated by solid stars, is characterized by a flat G$^{\prime}(\omega)$ and a weakly
frequency-dependent G$^{\prime\prime}(\omega)$. The magnitude of G$^{\prime\prime}$ is significantly larger than the expected response of the continuous fluid phase, but is always lower than the measured values of G$^{\prime}$ over the entire range of accessible angular frequencies. These features in the frequency response that correspond to samples in regime 2 (the example data is shown in inset (b) for a sample concentration = 0.25 g/c, acquired at a temperature of 60$^{\circ}$C) are typical of soft solid-like materials. The third regime, identified by open squares, exists for
the samples at the highest concentrations and at appropriately high temperatures. Inset (c) shows a frequency response measurement of a sample of concentration 0.29 g/cc, acquired at 60$^{\circ}$C, which lies in this regime. The frequency dependences of G$^{\prime}$ and G$^{\prime\prime}$ are similar to those measured in regime 2, but the magnitudes of the moduli are larger by almost factors of 100. We identify this third regime with the face-centred cubic phase
observed by Yardmici {\it et al.} in their SANS experiments \cite{yardimci_jcp}. We further identify regime 2, which is characterized by  G$^{\prime} >$ G$^{\prime\prime}$, but where the magnitudes of both moduli are significantly lower than those seen in the third regime, as a coexistence region between the micellar fluid and grains of solid crystalline phases, analogous to the phase observed in the diblock copolymer system E$_{87}$B$_{18}$
\cite{castelletto_jcp}. We should note here that similar frequency-dependence of the mechanical moduli observed in recent rheology studies performed by Saxena {\it et al.} on poly(diethylsiloxane) melts \cite{saxena_jrheol} have been attributed to the presence of mesophase domains that serve as
effective crosslinks in the melt. 

In a previous study by Lau et al. \cite{lau_jpol}, the authors use rheological measurements to arrive at a phase diagram of Synperonic F-108 solutions. These authors comment on a `micellization phase' that lies between the low-concentration, low-temperature regime (where the sample is expected to consist of unimers in solution) and the solid-like phase that exists at high temperatures and high concentrations (which the authors refer to as the `gel' phase). The authors do not comment on the specific nature of this micellization phase, but describe it as an `introductory transition regime from a micelle towards a gel'. Such an intermediate phase has not been reported in the neutron scattering study reported in \cite{yardimci_jcp}. Molino {\it et al.} \cite{molino_epjb} report an intermediate phase characterized by macroscopic segregation of the liquid like and crystalline phases, with eventual sedimentation of the solid like phase. We would like to point out here that we do not observe sedimentation in our samples that lie in regime 2 of our phase diagram, even after waiting for one week. For polycrystalline samples of concentrations 0.28 g/cc and below, and at the highest temperatures achieved in this study, the mechanical moduli decrease to lower values that we have previously identified with regime 2. This is believed to happen as at the highest temperatures, the deteriorating solvent conditions result in dehydration of the EO chains and the eventual collapse of the micellar polycrystals. For our samples at concentrations of 0.29 g/cc and above, this reentrance to regime 2 presumably occurs at temperatures higher than the maximum temperature of 80$^{\circ}$C achieved in this study. This feature is clear from our phase diagram (Fig. 3).

\subsection{Slow dynamics in F-108 samples: estimates from fits to frequency response measurements {\it vs.} results obtained from strain rate frequency superposition experiments}  

Fig 4 plots the frequency dependent elastic (solid squares) and viscous (solid circles) moduli of an F-108 sample of concentration 0.28 g/cc. For these measurements, the strain amplitude is kept fixed at 0.5\% and the sample temperature is maintained at 60$^{\circ}$C. Typically, frequency response measurements are characterized by a structural relaxation peak in G$^{\prime\prime}$, which, in the case of the Maxwell model for linear viscoelasticity \cite{maxwell_phil}, coincides with the crossover between the liquid-like and solid-like behaviors of the sample. The inverse of the angular frequency at which this relaxation peak occurs is identified with the characteristic relaxation time of the sample. Since our sample is a soft solid, we expect the dynamics to be slow. From fig. 4, we see that this structural relaxation peak is absent, presumably because it occurs at a very low frequency that lies outside our measurement window. Nevertheless, we attempt to extract a value for a characteristic time scale by modeling our data suitably: we attribute the slow dynamics in our sample to a broad band of solid-like properties and fit the data to the Cole-Davidson model, in which the complex modulus G$^{\star} = G_{\circ}[1-{\frac{1}{(1+i\omega\tau_{\circ})^{\alpha}}}]$, where G$_{\circ}$ is a high-frequency plateau modulus, $\tau_{\circ}$ is an estimate for an average relaxation time and the parameter $\alpha$ varies between 0 and 1 ($\alpha$ = 1 corresponds to the Maxwell model for linear viscoelasticity). This empirical model was first proposed to describe the dispersion and absorption of some liquids and dieletrics \cite{cole_jcp} and to model the broad spectrum of dielectric relaxation times observed in propylene glycol and {\it n}-propanol \cite{davidson_jcp}. In fig. 4, G$^{\prime}$ and G$^{\prime\prime}$ are fitted to the real and imaginary parts respectively of the Cole-Davidson model. These fits are indicated by solid lines in fig. 4 and give G$_{\circ}$ = 12400 $\pm$ 180 Pa, $\tau_{\circ}$ = 52 $\pm$ 16 secs and $\alpha$ = 0.37 $\pm$ 0.1.   

We next attempt to measure and understand the slow dynamics of this soft solid by conducting strain rate frequency superposition (SRFS) experiments \cite{wyss_prl,miyazaki_epl}, a technique proposed recently to access relaxation processes that are too slow to be measured in typical frequency sweep measurements. In contrast to our frequency sweep measurements where we extrapolated an average value for the characteristic relaxation time, in our SRFS experiments we simply apply strains that are large enough to speed up the dynamics of our soft solids and drive the structural relaxation peak of G$^{\prime\prime}$ to higher, and therefore, experimentally accessible frequencies. Since the time scale of the structural relaxation process is set by the rate of strain, the strain rate amplitudes are kept fixed in these measurements while the experiment sweeps over a range of frequencies. The time scale of the characteristic relaxation process $\tau(\dot\gamma)$ varies with the strain rate $\dot\gamma$ according to the empirical relation $\frac{1}{\tau(\dot\gamma)}$ = $\frac{1}{\tau_{\circ}}$+ K${\dot\gamma}^{\nu}$. Most metastable sytems with slow dynamics show K $\sim$ 1 and $\nu \sim$ 1 \cite{miyazaki_epl}. 

Fig. 5 shows the raw SRFS data for our 0.28 g/cc sample at 60$^{\circ}$C, where the strain rate amplitude is held fixed at 8 values between 0.0003/sec (denoted by down triangles), where the response is presumably linear, and 5/sec (denoted by stars) where the response is strongly nonlinear. In this figure, the data at each strain rate is designated by a particular symbol (see figure caption for details), with the solid symbols indicating G$^{\prime}$, and the corresponding open ones indicating G$^{\prime\prime}$. As the strain rate is increased, we see that the peak in G$^{\prime\prime}$ shifts monotonically to higher frequencies that eventually fall within our experimentally accessible frequency range. To achieve a particular strain rate amplitude $\dot\gamma$ in these experiments, we vary the strain amplitude $\gamma_{\circ}$ and the angular frequency $\omega$ such that $\dot\gamma = \gamma_{\circ}\omega$. That the strain amplitude varies inverse linearly with the applied angular frequency in all our SRFS experimental runs is shown in the inset of figure 5.  

To highlight the similarities in the shapes of the elastic and viscous moduli measured in these SRFS experiments, we follow the recipe suggested in \cite{wyss_prl} and rescale the moduli according to the relation G$^{\star}_{scaled}(\omega) =$ G$^{\star}(\omega/b(\dot\gamma))/a(\dot\gamma)$, where $a(\dot\gamma)$ and $b(\dot\gamma)$ are the scaling factors for the moduli and the angular frequencies respectively. Fig. 6 shows the master curve that we obtain by rescaling the data according to this recipe. The inset of this plot shows the dependence of the scaling factors $a(\dot\gamma)$ and $b(\dot\gamma)$ on the strain rate $\dot\gamma$. Here, $a(\dot\gamma) \approx$ 1 and is almost constant over the entire range of strain rates, while $b(\dot\gamma) \approx$ 1 for low strain rates (linear regime) and increases at higher strain rates (nonlinear regime) according to the power law $b(\dot\gamma) = {\dot\gamma}^{\nu}$, where $\nu \sim$ 0.86. This technique is reminiscent of the time-temperature superposition method frequently used in polymeric systems \cite{ferry_polymer}. Rescaling the data using this protocol, we extend the angular frequency range to much lower values and can therefore now easily isolate the structural relaxation peak. The peak in the rescaled G$^{\prime\prime}$ occurs at $\omega \sim$ 0.14 rad/sec, which indicates a structural relaxation time of $\tau_{\circ}$ = 45 $\pm$ 5 secs for this sample. This value, obtained using SRFS, compares very well with our previous estimate obtained by fitting the frequency response data to the Cole Davidson model (fig. 4), but is far more accurate than the relaxation time calculated from the fit. We believe that this time scale reflects the slow interfacial dynamics at the grain boundaries \cite{gottstein_gb,boyer_pre} between randomly oriented crystallites in this polycrystalline material. We have also performed systematic SRFS measurements to calculate the relaxation dynamics of polycrystalline F-108 samples of different concentrations, and find that the characteristic time scales measured exhibit a very strong dependence on the sample concentration \cite{harsha_unpublished}. This result complements nicely the results in \cite{lau_jpol} where the authors find that G$^{\prime}$ increases rapidly with sample concentration according to the relation G$^{\prime} \sim$ c$^{7.3}$.    

\section{Conclusions} 

We have performed a systematic study of the phase behavior and slow dynamics of solutions of the triblock copolymer F-108, using oscillatory rheology as the main experimental tool. In order to identify the linear viscoelastic range of these materials, we first study the response of F-108 solutions to oscillatory strains. For a range of sample concentrations and temperatures, no linear viscoelastic regime can be isolated. At higher values of oscillatory strains lying above a critical value, the moduli are found to decay as power laws. The data for G$^{\prime\prime}$ in these experiments are characterized by prominent peaks. These features in the oscillatory strain response of F-108 solutions are identified with soft solid-like behavior, ubiquitous in a diverse array of soft materials such as concentrated suspensions, pastes, emulsions and foams. Interestingly, the power laws characterizing the decays of the elastic and viscous moduli are insensitive to the sample concentration. We next construct a phase diagram of the sample in the concentration-temperature plane by performing temperature sweep experiments for samples of several concentrations. Abrupt changes in the magnitudes of G$^{\prime}$ and G$^{\prime\prime}$ with temperature are identified with transformations between different phases. We perform frequency response measurements to identify these phases and use this information to construct a concentration-temperature phase diagram for F-108. We show the existence of a new elastic phase where the moduli G$^{\prime}$ and G$^{\prime\prime}$ are much weaker than the values expected for a polycrystalline soft solid. This phase, previously not observed in triblock copolymer systems, is identified as a coexistence region between crystallites consisting of closely-packed micelles arranged on a cubic lattice, and a micellar fluid. As the temperature and concentration of F-108 is raised, these crystallites grow in size, which is reflected by the gradual increase in the values of the mechanical moduli. Finally, we measure the slow dynamical time scales of F-108 solutions by performing frequency sweep and SRFS experiments.  We note here that the scaling factor for the frequency $b(\dot\gamma)$, obtained by scaling the SRFS data onto a master plot, shows a power law dependence described by $b(\dot\gamma) = {\dot\gamma}^{\nu}$, where $\nu$ = 0.86, very close to the exponent of the power law decay of G$^{\prime\prime}$ with strain (G$^{\prime\prime} \sim \gamma_{\circ}^{-\nu^{\prime\prime}}$, where $\nu^{\prime\prime} \sim$ 0.82 $\pm$ 0.06). The measured relaxation time is identified with the slow, metastable dynamics at the interfaces between the polycrystalline domains and is found to depend strongly on sample concentration \cite{harsha_unpublished}. The metastability exhibited by these samples is further illustrated by our observation that the complex and shear viscosities of these materials do not satisfy the empirical Cox-Merz rule \cite{harsha_unpublished,sollich_pre}. Detailed experiments to understand the origins of the metastable dynamics in block copolymer solutions, and the effects of the chain lengths and molecular weights on their phase behavior and dynamics will form the subject of a future publication.

\newpage  

\newpage
\begin{figure}
\begin{center}
\includegraphics[width=3in]{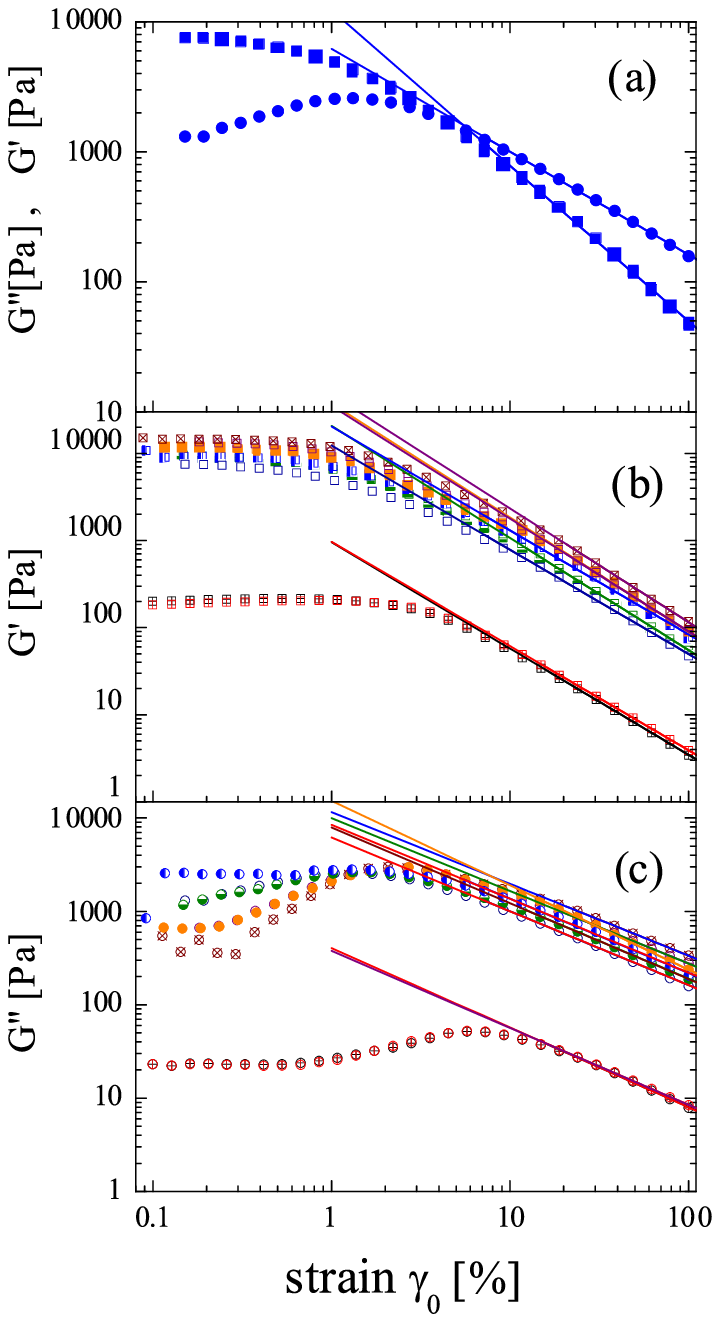}
\caption{(Color online) Amplitude sweep data for an F-108 solution of concentration 0.25 g/cc at a temperature of 60$^{\circ}$C is shown in (a). Above a certain critical strain, the moduli decay as power laws. The solid lines show the fits of G$^{\prime}$ (denoted by squares) and G$^{\prime\prime}$ (denoted by circles) to power laws with exponents 1.24 $\pm$ 0.06 and 0.82 $\pm$ 0.06 respectively. Amplitude sweep measurements and power-law fits to the data for G$^{\prime}$ for various sample concentrations ranging from 0.16 g/cc and 0.29 g/cc are shown in (b). Amplitude sweep measurements and power-law fits to the data for G$^{\prime\prime}$ for 6 sample concentrations between 0.16 g/cc and 0.29 g/cc are shown in (c). The exponents of the fits to the power-law decays do not change with changes in sample concentration.\\}
\label{Fig 1.}
\end{center}
\end{figure}
\newpage
\begin{figure}
\begin{center}
\includegraphics[width=3in]{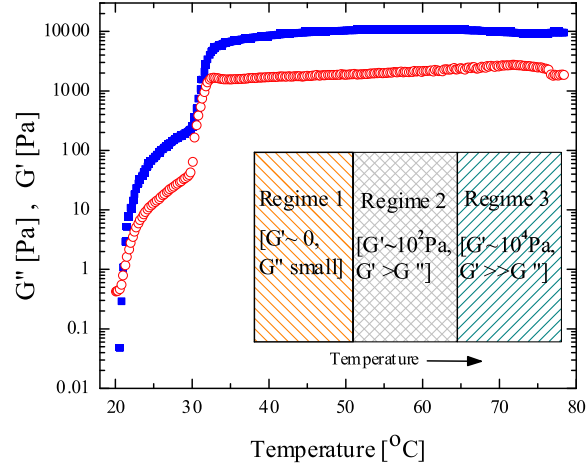}
\caption{(Color online) Temperature dependence of G$^{\prime}$ (solid squares) and G$^{\prime\prime}$ (open circles) for an F-108 solution of concentration 0.28 g/cc. At temperatures less than 20.1$^{\circ}$C, G$^{\prime} \sim$ 0 and G$^{\prime\prime} \sim$ 0.5 Pa. At approximately 20.2$^{\circ}$C, both moduli show a gradual increase. Around 30$^{\circ}$C, the moduli show a second, more rapid increase, before leveling off to a plateau characterized by G$^{\prime} \sim$ 10000 Pa and G$^{\prime\prime} \sim$ 1000 Pa. Depending on the absolute values of G$^{\prime}$ and G$^{\prime\prime }$ with increasing temperature, the data can be divided into 3 regimes (see inset).\\}
\label{Fig 2}
\end{center}
\end{figure}

\newpage
\begin{figure}
\begin{center}
\includegraphics[width=3in]{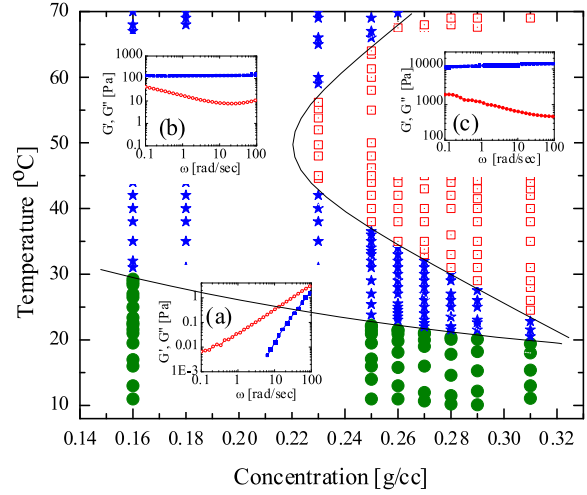}
\caption{(Color online) Phase diagram in the temperature-concentration plane of F-108 solutions. We identify the following three phases: a micellar fluid phase coexisting with dissolved unimers (solid circles), a soft solid phase (solid stars), possibly indicating a coexistence between randonly oriented crystallites and the micellar fluid, and a polycrystalline phase (open squares), characterized by very high values of the mechanical moduli. The phase boundaries are determined from our temperature sweep data, and the insets correspond to frequency response measurements in the different phases.\\}
\label{Fig 3}
\end{center}
\end{figure}

\newpage
\begin{figure}
\begin{center}
\includegraphics[width=3in]{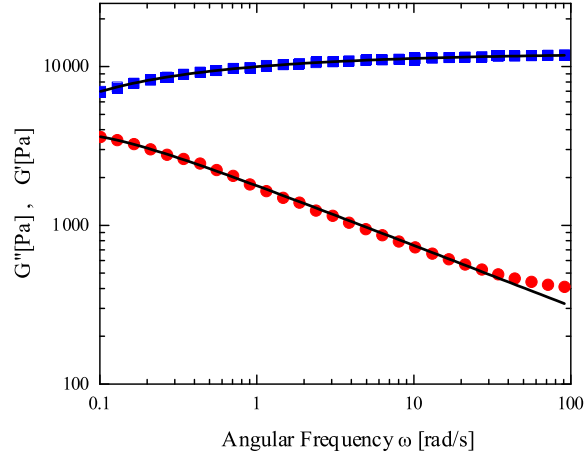}
\caption{(Color online) G$^{\prime}(\omega)$ (solid squares) and G$^{\prime\prime}(\omega)$ (solid circles) data {\it vs.} angular frequency $\omega$ for an F-108 solution of concentration 0.28 g/cc, at 60$^{\circ}$C. The solid lines through G$^{\prime}$ and G$^{\prime\prime}$ are fits to the real and imaginary parts respectively of the Cole-Davidson model.\\}
\label{Fig 2}
\end{center}
\end{figure}

\newpage
\begin{figure}
\begin{center}
\includegraphics[width=3in]{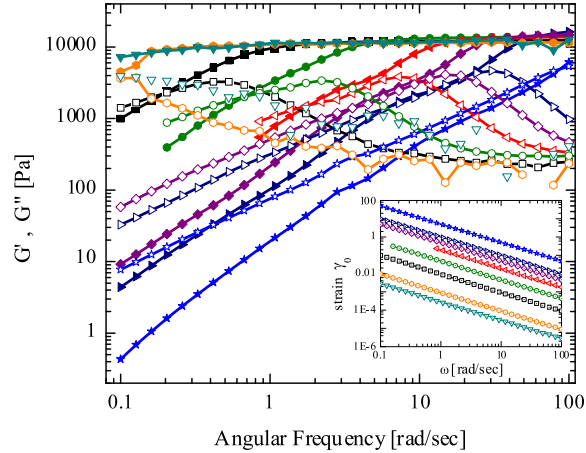}
\caption{(Color online) G$^{\prime}(\omega)$ (solid symbols) and G$^{\prime\prime}(\omega)$ (corresponding open symbols) {\it vs.} $\omega$ acquired at 8 constant strain rate amplitudes varying between 0.0003/sec (down triangles) and 5/sec (stars) for F-108 solutions of concentration 0.28 g/cc, at 60$^{\circ}$C.  The data for the intermediate shear rate amplitudes are denoted as follows: circles (correspond to a shear rate of 0.001/sec), squares (0.01/sec), up triangles (0.05/sec), left triangles (0.2/sec), diamonds (0.5/sec), and right triangles (1/sec). The inset shows the plots of the applied strain amplitudes $\gamma_{\circ}$ at each $\omega$ that yield constant strain rate amplitudes in our SRFS measurements. \\}
\label{Fig 5}
\end{center}
\end{figure}

\newpage
\begin{figure}
\begin{center}
\includegraphics[width=3in]{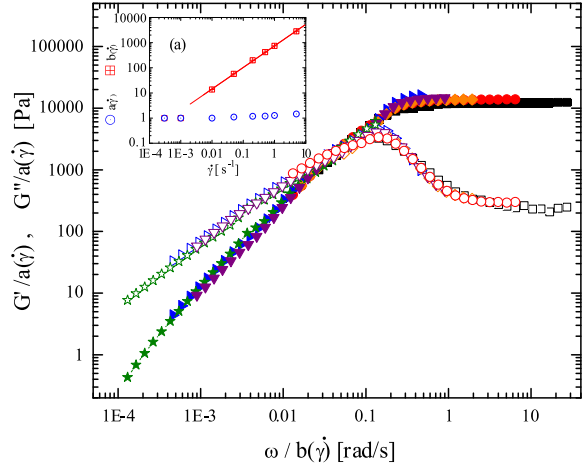}
\caption{(Color online) Master curve for the rescaled moduli G$^{\prime}/a(\dot\gamma)$ (solid symbols) and G$^{\prime\prime}/a(\dot\gamma)$ (open symbols) {\it vs.} the rescaled angular frequency $\omega/b(\dot\gamma)$, acquired for 8 shear rate amplitudes between 0.0003/sec and 5/sec. The inset shows the dependence of the scaling factors $a(\dot\gamma)$ and $b(\dot\gamma)$ on $\dot\gamma$.\\}
\label{Fig 6}
\end{center}
\end{figure}

\end{document}